\documentclass{appolb}
\usepackage{epsfig}

\begin{document}
\eqsec  
\title{Properties of Extensive Air Showers
\thanks{Invited talk given at the Epiphany Conference on Astroparticle
Physics, January 8-11, 2004, Krak\'ow, Poland.}
}
\author{Markus Risse
\thanks{Electronic address: markus.risse@ifj.edu.pl}
\address{
H.~Niewodnicza\'nski Institute of Nuclear Physics, Polish Academy of Sciences,
ul.~Radzikowskiego 152,
31-342 Krak\'ow, Poland
\\
Forschungszentrum Karlsruhe, Institut f\"ur Kernphysik,
 76021 Karlsruhe, Germany
}
}
\maketitle
\begin{abstract}
Some general properties of extensive air showers are discussed.
The main focus is put on the longitudinal development,
in particular the energy flow,
and on the lateral distribution of different air shower components.
The intention of the paper is to provide a basic introduction
to the subject rather than a comprehensive review. 
\end{abstract}
  
\section{Introduction}
Extensive air showers (EAS) are known for about 70 years to be
cascades initiated by primary cosmic rays in the atmosphere.
As such, they serve as connection to the highest particle energies
nature is offering, especially at energies exceeding $10^{15}$~eV
where direct measurements of cosmic rays are hampered by the low
primary flux. 
EAS can be viewed as tools for astroparticle physics.
The main task is to reconstruct from the shower
observables the parameters of the initial cosmic ray, which is
in general straightforward for the particle direction. It is more a challenge
for the energy and much more for the particle type of the primary.
This is due mostly to shower fluctuations, an important EAS property given
by the stochastic nature of the elementary interaction processes.
Combined with the steep primary spectrum, the
fluctuations impose a particular challenge for interpreting air shower data.
Use can be made, however, of the fact that an EAS consists of
different shower components.

The main body of this paper deals
with illustrating, supported by EAS simulations,
some characteristics of these shower components and the
information they contain about the primary particle.
This might serve as introduction to the subject of extensive air showers
and help to better understand what air shower experiments
are measuring, which information they try to infer from the data,
and where limitations are given~\cite{proc_exp}.
A more comprehensive introduction to EAS is given e.g.~in~\cite{gaisser}.

\begin{figure} \centering
  \epsfig{file=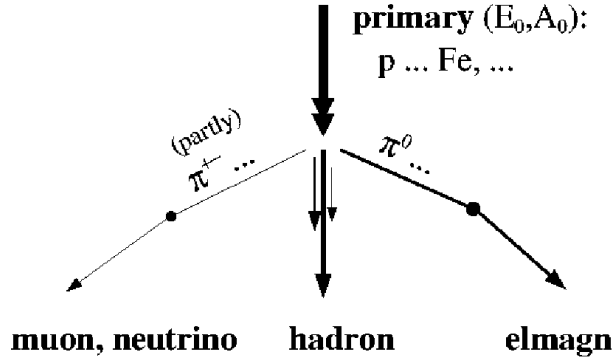,width=0.73\textwidth}
  \caption{Schematic sketch of an EAS.}
  \label{fig-scheme}
\end{figure}

A schematic sketch of an EAS is given in Figure~\ref{fig-scheme}.
The incident particle (primary energy $E_0$ and primary type $A_0$)
hits an air nucleus. Usually considered as extreme illustrations of
parimary particle types are protons and iron nuclei
(although an analysis of EAS data is not restricted to this range).
The notation $A_0$ anticipates the
fact that EAS features of primary hadrons, which are known as dominant
cosmic rays from direct measurements at energies below $10^{15}$~eV,
show some dependence on the nucleon number $A$ of a nucleus, as
explained later.
In the first interaction, a number of secondary hadronic particles
are generated. Mostly by decay of the neutral pions, electromagnetic
sub-cascades are initiated, and by decay of (part of) the charged
pions, muons and neutrinos are produced. While neutrinos are hardly
detectable in classic air shower experiments, muons might reach the
observation level even from large altitudes. It is important
to note that a competition of the charged
pions between decay and interaction exists which depends on the
pion energy and the local air density. Some information on the first
interactions will be imprinted in this way in the muon component.
The remainder of the hadronic secondaries continues to interact with
air in subsequent collisions and to feed the other shower components.

In ground arrays, the surviving particles are measured with 
particle detectors,
and with appropriate optical telescopes Cherenkov and fluorescence
emission during the shower propagation can be observed.

To characterize more quantitatively the interaction process, it is
helpful to introduce the quantity {\it inelasticity k} as the energy fraction
available for the production of secondary particles (or in other words,
the initial collision energy reduced by the energy of the most
energetic particle). As a rule of thumb, about $\frac{1}{3} k$ is ``lost''
to the electromagnetic channel per hadronic interaction. 
For a mean value $k \simeq 0.6$ typical for high-energy interactions
this corresponds to $\simeq 20\%$ per interaction that on average will
give rise to subsequent electromagnetic cascading.
Equipped with this knowledge, we can turn to the longitudinal shower
development.

\section{Longitudinal shower development}

\subsection{Energy flow}
\begin{figure}[t] \centering
  \epsfig{file=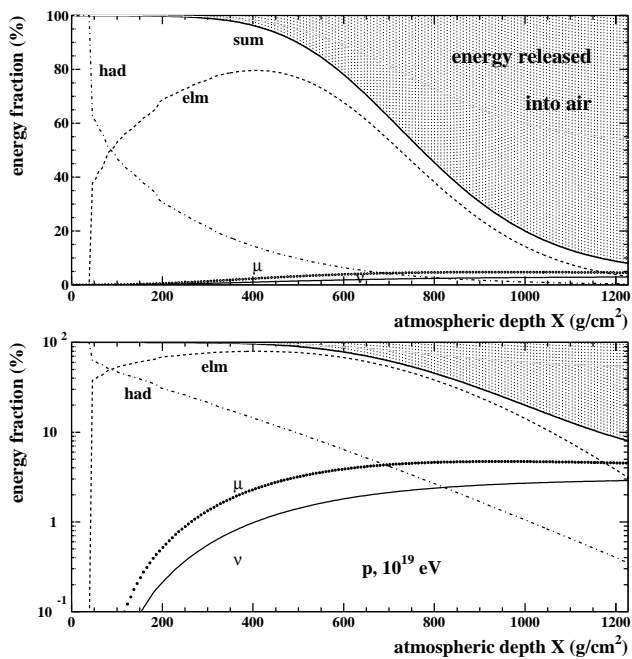,width=0.73\textwidth}
\caption{Energy flow in EAS as obtained by
CORSIKA shower simulations for an individual $10^{19}$~eV primary proton
event.
The energy fractions stored in hadrons, electromagnetic particles,
muons, and neutrinos are shown.
The difference between their sum to the initial energy
indicates the total amount of energy already released into air
(shaded area).
Upper graph in linear, lower graph in logarithmic scale.}
  \label{fig-flow1}
\end{figure}

At first the general energy flow in an air shower is discussed
in some detail.
In Figure~\ref{fig-flow1}, the energy carried by different shower
components during the cascade development are displayed for the
example of a proton-induced event of primary energy $10^{19}$~eV.
These are results of detailed Monte Carlo air shower simulations
obtained with the program package CORSIKA~\cite{corsika}.
For more information on simulation in particular at the highest energies,
see for instance~\cite{knapp} and references given therein.

Initially, all energy is concentrated in the primary proton.
For this event, the first interaction occurs after traversing an
atmospheric depth of $\simeq$40~g~cm$^{-2}$. A significant energy fraction
is transferred to the electromagnetic component, which is further fed 
in the subsequent shower process.
One can notice an exponential decrease of the energy left in
the hadrons. This can be understood in a simple picture of a constant
inelasticity $k$ and a constant mean free path $\lambda_h$ for hadronic
interactions. 
Based on the average fraction of $\frac{1}{3}k$ that is put into the
electromagnetic channel per hadronic interaction,
the energy remaining in the hadronic component
at depth $X$ is
\begin{equation}
\label{eq-hadenergy}
E_{h}(X) = E_0 \cdot (1-\frac{1}{3}k)^{X/\lambda_h} ~~.
\end{equation}
Adopting typical values used for modeling nucleon-air interactions at
high energies of $\lambda_h \simeq 55$~g~cm$^{-2}$ and $k \simeq 0.6 $,
the hadronic scale depth $\Lambda_h$ of the exponential fall-off
thus amounts to
\begin{equation}
\label{eq-lambda}
\Lambda_h = \frac{\lambda_h}{|\ln(1-\frac{1}{3}k)|} \simeq 
                                 250\mbox{ g cm$^{-2}$}
\end{equation}
in reasonable agreement with the results plotted for the detailed simulation.

In this approximation, energy losses into the muon and 
neutrino channels were assumed to be small. In fact, as can be
seen only at later stages of the shower development, an energy fraction of
around 5\% of the initial energy has accumulated in these components.
The different character of the longitudinal curves of these two components
with respect to hadrons and electromagnetic particles is evident.
It is an important feature of muons to transport most of their
(though small) energy fraction down to observation level.

\begin{figure} \centering
  \epsfig{file=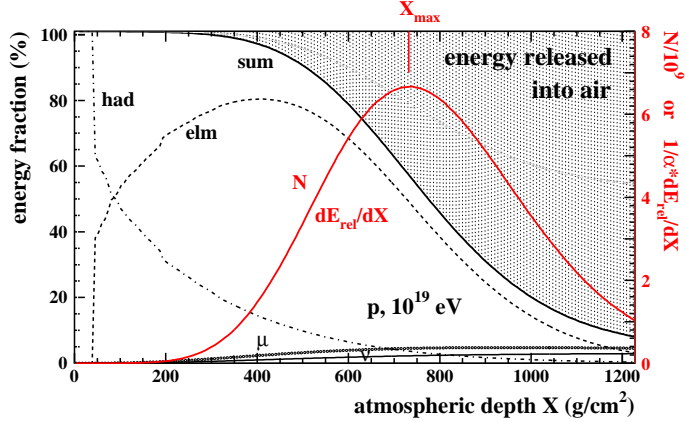,width=0.73\textwidth}
  \caption{Energy flow in EAS (left scale, compare Fig.~\ref{fig-flow1})
   and cascade profiles $N(X)$ or $1/\alpha\cdot dE_{rel}(X)/dX$ (right scale).
   The depth where
   the shower reaches its maximum number of particles is indicated
   by $X_{max}$.}
  \label{fig-flow2}
\end{figure}

The electromagnetic particle channel is rising fast. The process to extract
energy from the electromagnetic particles is the energy loss due to
ionization of air induced by the charged shower particles~\cite{risse}.
Therefore, in Figure~\ref{fig-flow2} the number of shower electrons
(a term which usually includes positrons) is overlayed to the
previous energy flow plot.
Since the specific ionization loss $\alpha$ of relativistic electrons
is about constant ($\alpha \simeq 2$~MeV/(g cm$^{-2}$)), the same curve
in appropriate units also resembles the differential energy release
$dE_{rel}(X)/dX$ of the EAS as a whole,
\begin{equation}
\label{eq-dedx}
\frac{dE_{rel}}{dX} (X) \simeq \alpha N(X) ~~.
\end{equation}
Integrating $dE_{rel}(X)/dX$ results in the
energy fraction indicated by the shaded region; for instance, at
shower maximum (labeled $X_{max}$), already $\simeq 50\%$ of the
initial energy has been released into the atmosphere.

The maximum of energy stored in 
electromagnetic particles $X_{elm} \simeq 410$~g~cm$^{-2}$
is reached well before the so-called 
shower maximum $X_{max} \simeq 730$~g cm$^{-2}$,
i.e.~the depth where the shower contains the
largest electron multiplicity.
This is due to the fact that at early cascade stages, a large energy fraction
is carried by high-energy particles. Only gradually, the energy
is transformed to newly created particles.
The maximum of energy stored in electromagnetic particles $X_{elm}$
is expected at the development stage
where the gain from the hadron
channel equals the loss by energy release,
\begin{equation}
\label{eq-xelm1}
-\frac{dE_{had}}{dX} (X_{elm}) \simeq \frac{dE_{rel}}{dX} (X_{elm}) ~~.
\end{equation}

We can roughly cross-check $X_{elm}$ in our simplified approach.
Using Eq.~(\ref{eq-dedx}) and (by differentiating Eq.~(\ref{eq-hadenergy}))
$-dE_h(X)/dX \simeq E_h(X)/\Lambda_h$
we obtain from Eq.~(\ref{eq-xelm1}) for $X_{elm}$ the condition
\begin{equation}
\label{eq-xelm2}
E_h(X_{elm}) \simeq \alpha \Lambda_h N(X_{elm}) ~~.
\end{equation}
Adopting (rounded) values yields an expectation of
$E_h$(410~g~cm$^{-2}) \simeq 2\cdot250\cdot2\cdot10^9$~MeV = $10\% \cdot E_0$
which is in reasonable agreement with the data.

Equations (\ref{eq-xelm1}) and (\ref{eq-xelm2}) indicate the connection
between energy flow (of hadronic to electromagnetic channel)
and particle multiplication (in the electromagnetic channel).
Most important for the observation are the profiles $dE_{rel}(X)/dX$
or $N(X)$, respectively. These profiles are observable as fluorescence light
and indicate how many particles reach the observation level.
We therefore next investigate some characteristics of cascade curves formed
by particle multiplication.

\subsection{Profile features}
Some of the main features of shower profiles can be nicely motivated
within a simple toy model of particle cascades~\cite{heitler},
which complements to the previous toy model of energy flow.
Let us suppose a particle with energy $E_0$ that splits its
energy equally into two particles after traveling a path-length $\lambda$,
and let this process be repeated by the secondaries.
We then obtain a particle cascade which at a path-length $X$ has evolved
into
$N(X) = 2^{X/\lambda}$
particles of equal energy
$E(X) = E_0/N(X)$.
Let us further assume that particle multiplication stops when a
certain energy limit $E = E_l$ is reached. Then, the maximum number of
particles $N_{max}$ is reached at this point $X_{max}$, and it is given by 
$N_{max} = N(X_{max}) = E_0/E_l$.
The position of $X_{max}$ follows as
$X_{max} = \lambda/\ln2 \cdot \ln(E_0/E_l)$.

\begin{figure} \centering
  \epsfig{file=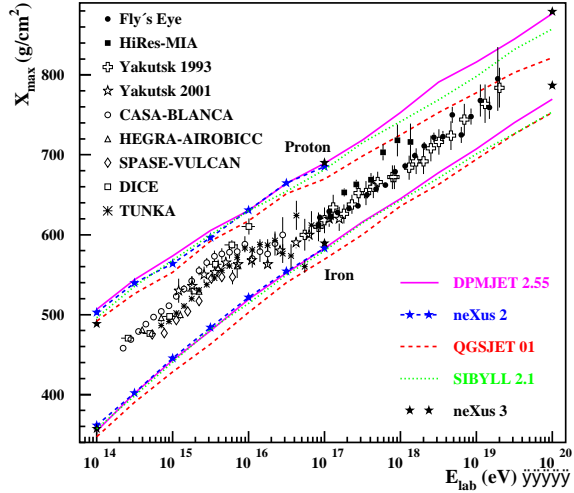,width=0.60\textwidth}
  \caption{Average depth of shower maximum as a function of
           primary energy~\cite{xmax-heck}.}
  \label{fig-xmax}
\end{figure}

\begin{figure} \centering
  \epsfig{file=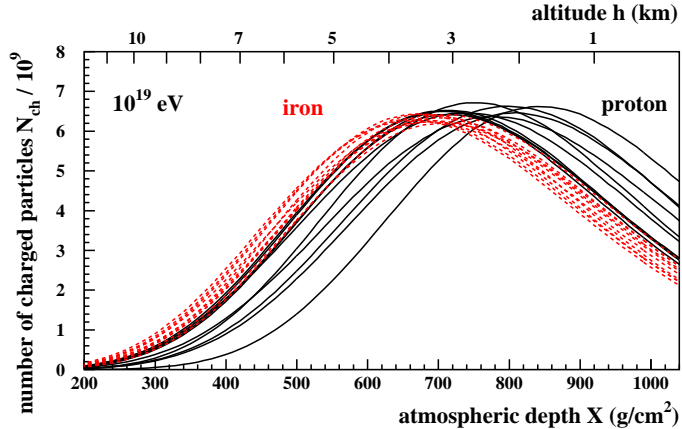,width=0.73\textwidth}
  \caption{Individual longitudinal shower profiles (vertical incidence).} 
  \label{fig-longifluct}
\end{figure}

Within this toy model,
let us construct the more general case of 
an initial set $(A_0, E_0/A_0)$ of $A_0$ particles, each with energy
$E_0/A_0$.
(The previous case is then realized for $A_0 = 1$.)
The particle number at maximum and the position of shower maximum
are then given by
\begin{equation}
\label{eq-toyn}
N_{max}(A_0, E_0/A_0) = N_{max}(E_0) \propto E_0 ~,
\end{equation}
\begin{equation}
\label{eq-toyx}
X_{max}(A_0, E_0/A_0) \propto \ln(E_0/ A_0 E_l) \le X_{max}(E_0) ~.
\end{equation}
If we identify the initial particle set $(A_0, E_0/A_0)$ as 
primary nucleus of mass number $A_0$ (this is known as ``superposition
model''), the toy model predictions can thus be summarized as
\begin{itemize}
\item $N_{max}$ increases proportional to the primary energy
\item $N_{max}$ is the same for all nuclei
\item $X_{max}$ increases with the logarithm of the primary energy
\item $X_{max}$ is smaller for the heavier nuclei (logarithmic
                dependence on $A_0$)
\item $X_{max}$ is the same for same $E_0/ A_0$ but different $E_0$
\end{itemize}
Despite the obvious simplicity (and its limitations) of this toy model,
the findings of detailed shower simulations are quite well reproduced,
as illustrated in Figures~\ref{fig-xmax} and \ref{fig-longifluct}:
The average $X_{max}$ increases with $\ln E_0$ and is smaller
for primary iron ($A_0$ = 56) compared to primary proton, with a difference
of $\simeq$~80$-$100~g~cm$^{-2}$
(Fig.~\ref{fig-xmax}).
The primary iron $X_{max}$ for $E_0$ fits well to the proton one of
energy $E_0 / 56$ (same $E_0/ A_0$).
$N_{max}$ is quite similar for the
different primaries (Fig.~\ref{fig-longifluct}) and about proportional
to $E_0$ (not shown).

Two other important EAS properties should be pointed out from 
Fig.~\ref{fig-longifluct}. Firstly, for fixed $E_0$ 
different $X_{max}$ values
translate into different particle numbers on ground. For instance,
the primary proton showers result on average in a larger number of
ground particles compared to iron showers
if the observation level is beyond the maximum.
Thus, observation of $X_{max}$ or related quantities
provide information on the primary particle type.
Secondly, however, shower-to-shower fluctuations are visible,
that lead to partly overlapping distributions of
shower observables. As an example, the RMS($X_{max}$) of primary protons
is $\simeq 60$~g~cm$^{-2}$ and thus nearly as large as the difference
between the average values of proton and iron. This limits an
event-by-event assignement of a primary particle type, 
and composition analyses are usually performed with large event samples.
However, it can also be seen that iron shower fluctuations are smaller
(e.g.~RMS($X_{max}$) $\simeq 20$~g~cm$^{-2}$)
than the proton ones.
This might be used in turn to conclude about the primary composition
by studying the very fluctuations of shower observables
in a given event sample.


\subsection{Shower electrons and muons}

\begin{figure} \centering
  \epsfig{file=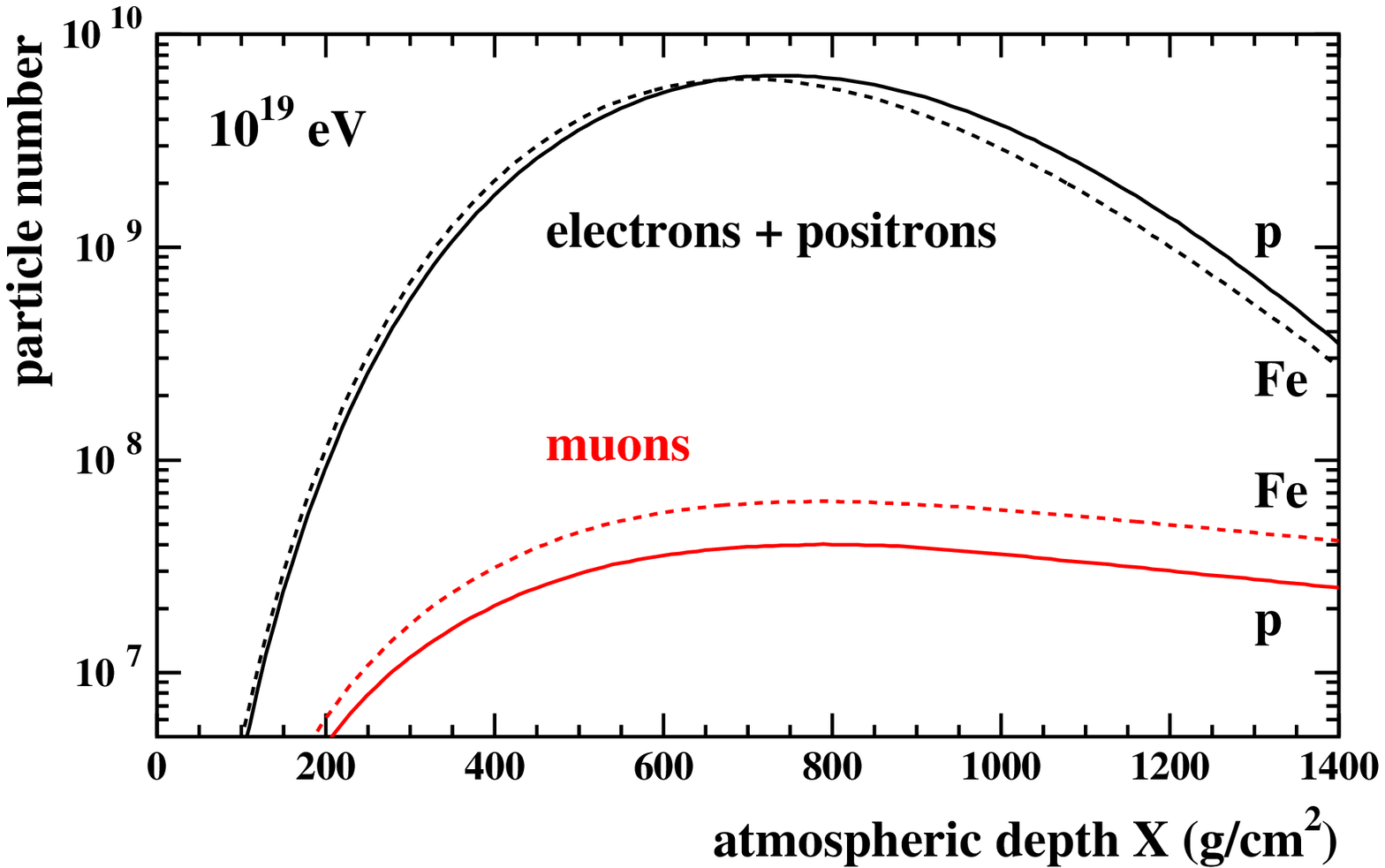,width=0.53\textwidth}
  \epsfig{file=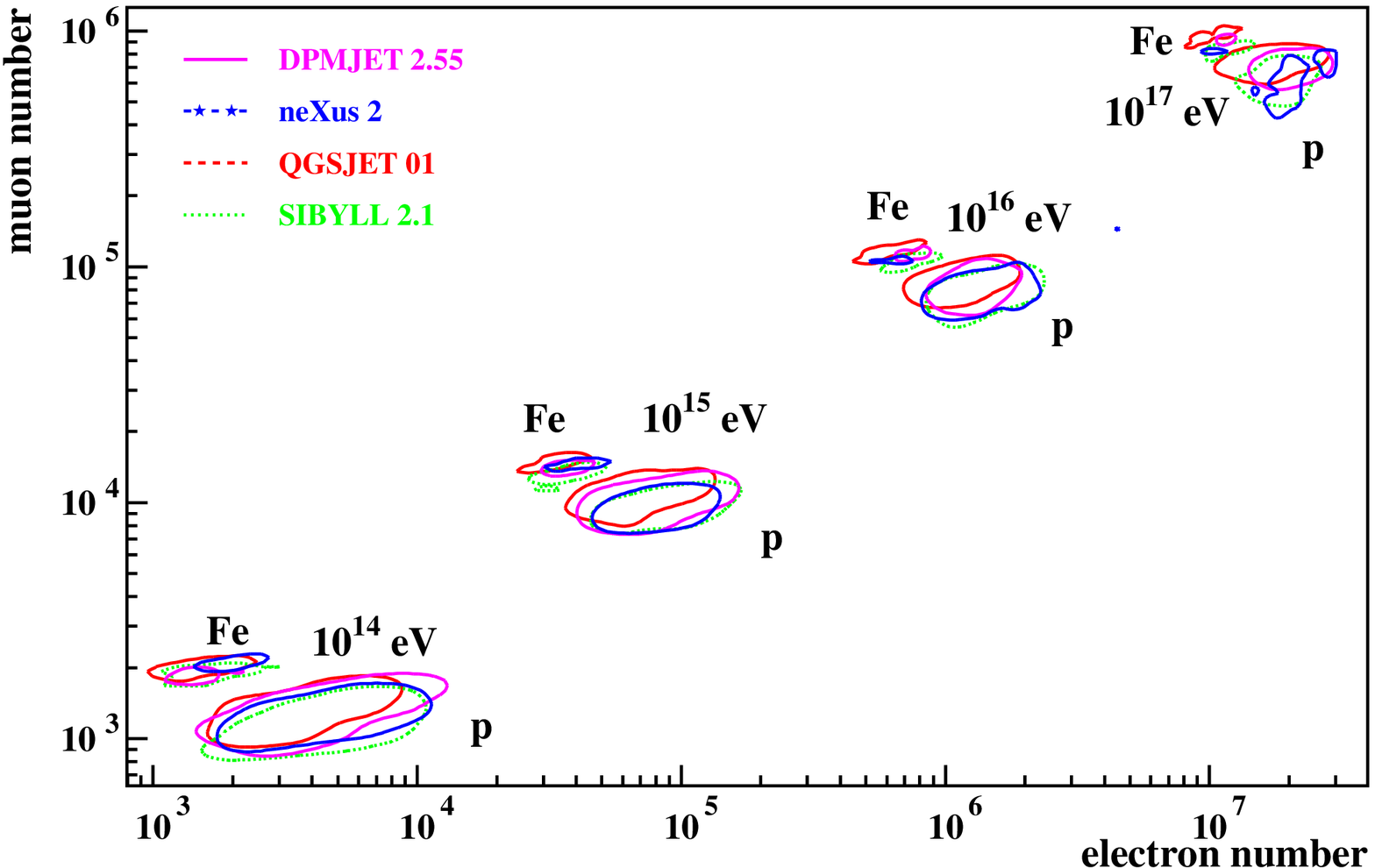,width=0.46\textwidth}
  \caption{({\it Left}) Longitudinal profiles of typical primary proton
   and iron events for shower muons and electrons.
   ({\it Right}) Distribution of total muon and electron number on ground
    for proton and iron induced showers of different primary
    energies~\cite{potato-heck}.}
  \label{fig-longiemu}
\end{figure}

The total number of shower particles regarded so far was dominated by the
shower electrons.
The shower muons, however, provide complementary information on the
primary particle. As visible from the longitudinal distributions
in Figure~\ref{fig-longiemu} (left),
the muon particle number is decreasing only slowly after the maximum,
in contrast to the shower electrons.
Moreover, the total muon number also depends on
the primary particle type. In the superposition model, the larger total muon
content in iron showers might be qualitatively understandable: Due to the
smaller energy per nucleon $(E_0/A_0)$, the secondary pions are less
energetic. This favours a pion decay as well as the fact that iron
events develop at larger altitudes, where the air density is smaller.

Shown in Figure~\ref{fig-longiemu} (right) are muon versus electron
number for
proton and iron induced events for different fixed primary energies.
The size of each ``potato'' corresponds to the shower fluctuation,
while the separation indicates
that a correlated measurement of ground particle numbers of shower
electrons and muons allows conclusions on the primary mass.

\section{Lateral distribution}

\begin{figure} \centering
  \epsfig{file=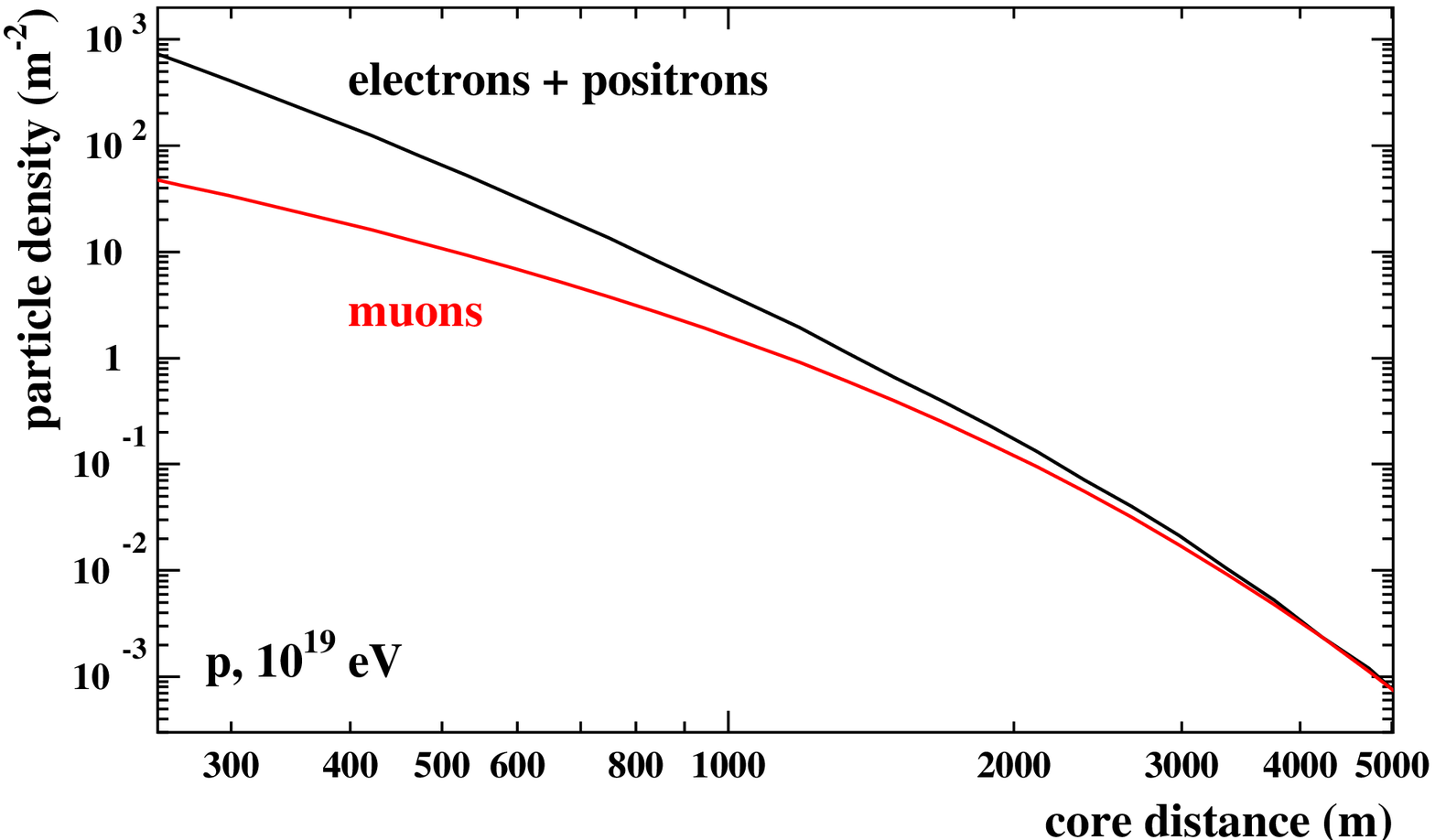,width=0.49\textwidth}
  \epsfig{file=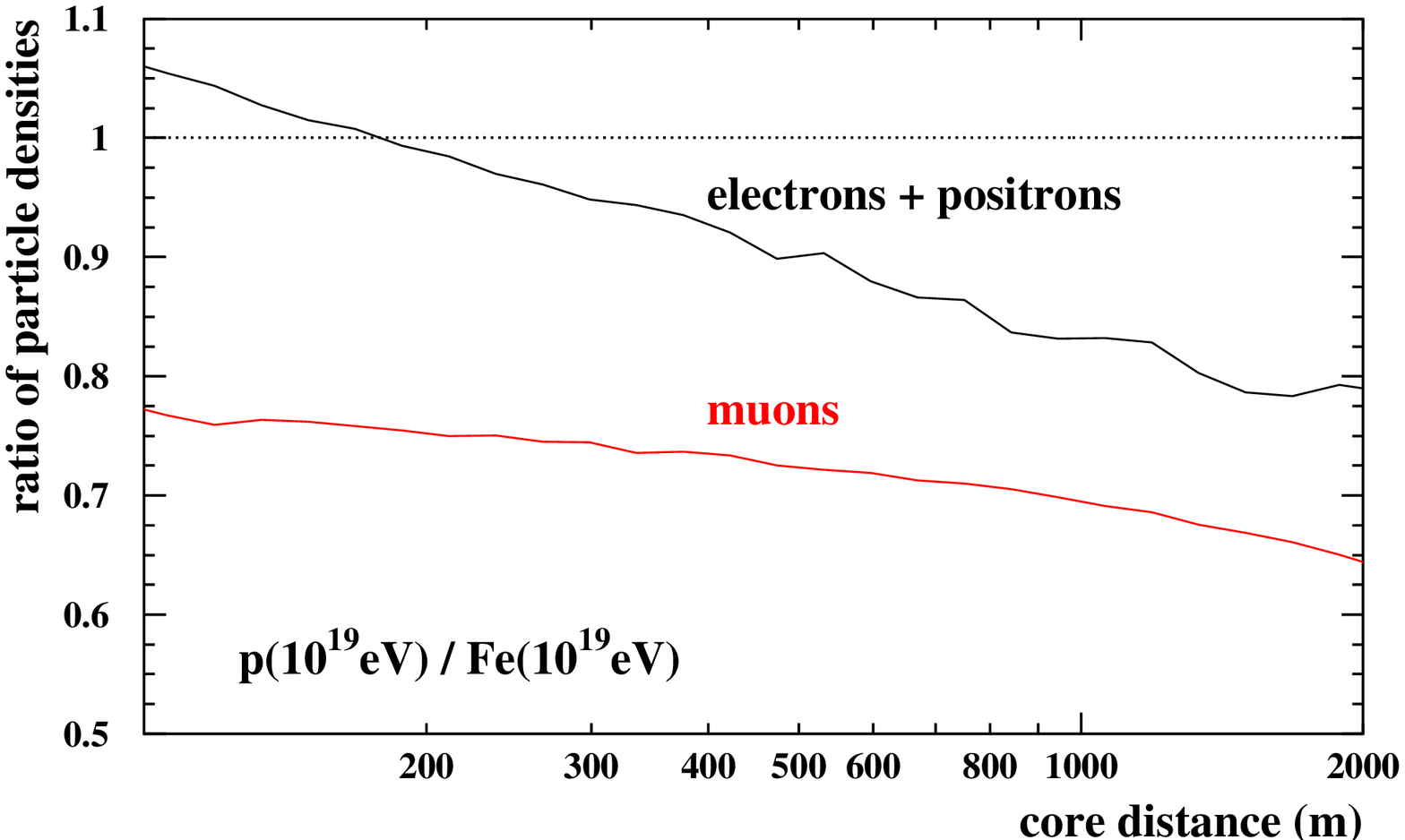,  width=0.49\textwidth}
  \caption{{\it (Left)} 
          Lateral distribution on ground of shower muons and electrons.
          {\it (Right)}
          Ratio of lateral distributions on ground in proton to iron induced
           events, both for shower muons and electrons.}
  \label{fig-lat}
\end{figure}

Only the longitudinal development was discussed so far. 
Air showers have a lateral spread that also differs for the different
shower components as well as for the various primary particles.
It can be seen in Figure~\ref{fig-lat} (left) that the lateral distribution
of shower muons on ground is flatter than the distribution of shower electrons.
This is mostly due to the muon origin from larger altitudes
compared to the more local production and fading of the electron
component. In spite of the electrons being much more numerous than
muons (about two orders of magnitude around shower maximum,
see Fig.~\ref{fig-longiemu}), at
larger distance from the shower core the particle densities become
comparable.

Given the differences in the longitudinal development between primary
proton and iron events, differences also in the lateral distributions
might be expected. In Figure~\ref{fig-lat} (right), the ratio of particle
densities of proton-induced to iron-induced events is displayed
for shower muons and electrons. Both ratios decrease with increasing
distance from the shower centre, indicating the flatter lateral
distributions in case of the (on average) more developed primary iron
showers. The larger muon content of iron-induced events is also visible
(ratio $< 1$). The larger electron number on ground for primary proton
showers can now be specified as larger ground particle density closer to
the core, while at larger distances (which contribute less to the
integrated, total particle number), the electron density in iron
events is larger due to the flat lateral distribution.
Thus, measuring local particle densities of the different shower components
as a function of core distance provides additional information on the
primary particle, which is utilized in large arrays of ground particle
detectors.

\section{Discussion and conclusion}

Extensive air showers consist of different particle components which
have different shower characteristics. This gives a handle to
determine primary energy and (to some extent) the primary particle type.
Extensions of this approach that were not discussed comprise
the exploitation of different time structures of the shower front
or comparing inclined events (where mostly muons survive the
increased atmospheric path-length) to near-vertical ones.
Also in the hadronic shower component, information on the
primary particle type is
imprinted; but maybe even more important is the possibility to test
high-energy hadronic interaction models used in the simulations.
Shower simulations have grown to an important tool for EAS data
reconstruction. Shower fluctuations motivate the
development of Monte Carlo simulation techniques.
In turn, these simulations might also be applied to find observables that 
depend less strongly on the specific primary or on
shower fluctuations, such as the signal at distances of 600$-$1000~m
from the core for primary energy estimations with ground arrays.
At the highest energies, further interaction features have to be
considered in shower simulations, e.g.~for primary photons the
LPM effect and preshower formation in the geomagnetic field.

In summary, even after many years of extensive exploration, air
showers continue to be fascinating tools for astroparticle physics.

{\it Acknowledgements.}
The conference organizers are thanked for their kind hospitality and
for creating an inspiring conference atmosphere. In particular, I would
like to thank Marek Je$\dot{\textnormal{z}}$abek
and Henryk Wilczy\'nski $-$ dzi\c{e}kuj\c{e} bardzo!
The fruitful collaboration with Ralph Engel and Dieter Heck is gratefully
acknowledged.
The author is supported by the Alexander von Humboldt Foundation.


\end{document}